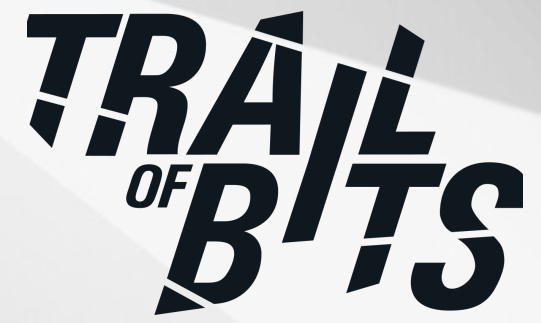

# Toward Comprehensive Risk Assessments and Assurance of AI-Based Systems

Heidy Khlaaf

**March 7, 2023**

**Recommended Citation:**

Khlaaf, Heidy. Toward Comprehensive Risk Assessments and Assurance of AI-Based Systems, Trail of Bits, 2023.

# Table of Contents

# Executive Summary

Public adoption and use of Artificial Intelligence (AI)-based systems have peaked in recent months due to the introduction of highly accessible AI tools and systems and the commercial trial of general multi-modal models such as GPT-3, Claude, LaMDA, Bard, and Stable Diffusion. Although AI was initially swamped with breathless marketing claims, its less glamorous potential harms—in the form of menacing or ethically questionable screeds emerging from these systems—have begun to emerge. With creators in many domains looking to quickly leverage this technology, consumers and the general public are at a loss as to how they are impacted, and whether these systems are safe and production ready.

Novel safety, socio-economic, and ethical harms arising from the deployment of AI-based systems have led to a breadth of work seeking to map, measure, and mitigate against newly found risks. These works have heavily leveraged techniques and terminology from the fields of System Safety Engineering and Cybersecurity, yet they have fallen short in accounting for the limitations and nuances that reduce the efficacy and correct application of adopted methodologies. Furthermore, misuse of terminology entailing compliance with established safety and security properties can mislead stakeholders with regard to the claims an AI system satisfies and provide a false sense of safety.

In this paper, we seek to align overlapping, AI-adjacent communities on a consistent and comprehensive assurance terminology crucial for the safe deployment of AI-based systems. We outline why previous attempts to adapt risk assessment techniques and terminology from the safety and security fields have been insufficient. We then propose a novel end-to-end AI risk framework that integrates the concept of an Operational Design Domains (ODD), initially introduced for ADS (Automated Driving Systems) [1], for more general AI-based systems. The purpose of an ODD is to provide a description of the specific operating conditions for which an AI-system is designed to properly behave, thus outlining the safety envelope for which system hazards and harms can be determined against. We believe that by defining a more concrete operational envelope, developers and auditors can better assess potential risks and required safety mitigations for AI-based systems.

This position paper seeks to be of interest to people in several broad groups. This paper is of most interest to those seeking to assure AI systems such as AI engineers, AI policymakers, AI auditors, and those intending to safely integrate AI systems into their operations or products. More broadly, we hope it appeals to members of the public who wish to understand the prospects of AI assurance and safety in the midst of marketing hype and exaggerated commercial messaging. Finally, we hope it informs those who may work in adjacent fields, who want to understand the complex trends of AI assurance and safety.

Overall, this paper provides the following key points and takeaways:

1. ***Distinguishing Safety and Alignment***. The AI community, conflating requirements engineering with safety measures, has allowed those building AI systems to abdicate safety by *equating safety measures with a system meeting its intent* (i.e., value alignment). Yet, in system safety engineering, safety must center on the lack of harm to others that may arise due to the system intent itself.

2. ***Limitations of Hardware Methodologies.*** Risk assessment techniques adopted from hardware safety (e.g., Failure Modes and Effects Analysis, Fault Tree Analysis) in the AI community are not suitable for AI-based systems. Attempting to measure safety properties of these systems through techniques developed under the assumption of random failures is not conducive to uncovering the design issues that directly lead to their systematic failures.

3. ***Scope of Safety vs. Security.*** The aim of safety is to prevent a system from impacting its environment in an undesirable or harmful way, typically to protect human lives, the natural environment, or assets. The aim of security, on the other hand, is to prevent often-adversarial environmental agents or conditions from impacting a system in an undesirable or harmful way. Therefore, safety risk frameworks may be more appropriate for exploring harms posed by a system (e.g., machine learning [ML] models) compared to threat modeling, which aims to protect a system from its external environment.

4. ***Use of System Safety Engineering.*** We recommend the use of more relevant system-level risk assessment frameworks such as MIL-STD-882e to build AI-specific risk frameworks on (see 5). More general systems engineering and software risk assessment frameworks have an expanded scope that aims to address hazards, harms, and systematic considerations crucial to software or AI, such as general system failures and emergent behaviors.

5. ***Assessing Safety through Operational Design Domains.***
    a. The majority of algorithmic assessments aim to audit general properties of a system without considering its operational envelope. The lack of a defined operational envelope for the deployment for general multi-modal models has rendered the evaluation of their risk and safety intractable, due to the sheer number of applications and, therefore, risks posed.
    b. We propose the integration of ODDs into a risk framework, where we define a novel ODD taxonomy relevant to the use of AI technologies, including general multi-modal models. The use of ODDs can guide in understanding the constraints under which the AI system may no longer behave as intended or how it can escape its designated safety envelope.

# 1. Introduction

Despite recent demonstrations of ever-increasing performance and ability across general domains, ML (Machine Learning) models have not only demonstrated a lack of robustness[1] in their outputs, but their performance and capabilities have proved difficult to measure [15, 27, 33]. Furthermore, the lack of clearly defined requirements and adequate risk analyses has led to safety hazards and novel socio-economic and ethical harms accompanying the deployment of Artificial Intelligence (AI)-based systems [4, 10, 27]. Yet few actionable or sufficient methodologies and mitigations have been defined to systematically address these hazards and harms, largely due to the fact that proposed AI risk frameworks and appropriate evaluation metrics have been insufficient in quantifying and qualifying novel AI failure modes and harms.

Frameworks aside, a lack of cohesion on baseline terminology, such as the distinction between “safety” and “alignment,” has led to contradictory approaches that improperly equate safety measures with a system meeting its intent, which can result in drastically different outcomes when constructing risk and hazard assessments. Indeed, despite AI-related works adapting and citing well-established system safety and security techniques, safety and risk terminologies have diverged from their established use and meaning. Therefore, In Section 2, we aim to align overlapping and AI-adjacent communities toward consistent and comprehensive terminology, which is crucial for the safe deployment of AI-based systems.

In Section 3, we outline how previous attempts to adapt hardware, cybersecurity, and system safety risk and safety techniques have been insufficient, and discuss nuances regarding fundamental issues that limit the direct application of said techniques to AI risk assessments. Subsequently, in Section 4, we propose a novel assurance approach that outlines a comprehensive risk assessment framework that overcomes many of the discussed limitations. In particular, the lack of a defined operational envelope for general multi-modal models has rendered the evaluation of their risk and safety intractable, due to the sheer number of risks posed. We propose the use of Operational Design Domains [1] (ODDs) within AI risk assessments, where we define a novel AI-based ODD taxonomy to allow for the exploration of a wide range of scenarios and their associated risks. The consideration of the ODD categories and their interactions helps operationalize risks under a selected application domain; aiming to help developers and auditors build confidence that an AI-based system has addressed its safety risks.

[1] It has been argued that a lack of robustness may be an inherent property to the way in which current ML models are constructed (e.g., Deep Neural Networks) [37]

# 2. Distinguishing Value Alignment, Safety, and Risk

Despite AI-related works adapting and citing well-established system safety and security techniques, various terminology has been misconstrued in its use and meaning. In this section, we seek to align the community on the terminology used, as a lack of consistent or intentional definitions compromises the integrity of the safety and security techniques the ML community seeks to adapt. Furthermore, the lack of alignment on the quantification of safety prevents the systematic execution of safety protocols necessary for preventing identified [4, 6, 44] and novel harms produced by AI-based systems.

## 2.1 Conflating 'Value Alignment' and 'Safety'

The term "safety" has come to have a multitude of definitions within AI, which vary based on the context and the community. These definitions have not fully captured the broader meaning of "safety" used within the fields of Systems and Safety Engineering, and may in fact be a direct contradiction to it. Within the context of AI communities, some have defined "safety" as the prevention of failures due to accidents [3, 35], while others refer to the field of Alignment, aiming to steer AI systems toward human-oriented values and goals [23, 8]. Not only are Alignment measures subjective at best, but they fundamentally conflate *safety properties* with *system requirements*, which are well-established engineering concepts. Compare the following established definitions:

- **Value Alignment** [8]: AI systems should be designed so that their goals and behaviors can be assured to align with human values throughout their operation.
- **System Requirement**: A statement that translates or expresses functionality to satisfy intent or stakeholders' needs.

Given that intent and stakeholders' needs are subjective human values, the term "Value Alignment" is a specific type of a system requirement. This confusion of terminology is not trivial, but rather critical, considering the definition of "safety" often deployed for safety-critical systems:

- **Safety**: To prevent a system from impacting its environment in an undesirable or harmful way, typically aiming to protect human lives, natural environment, or monetary assets.

That is, safety concerns are derived based on hazards or harms posed by a system meeting *its specifications*, where safety functionality is introduced to reduce the frequency (or probability) of the hazardous event or harms and/or its consequences that may occur. Conflating requirements engineering with safety measures has allowed those building AI systems to abdicate safety by *equating safety measures with a system meeting its intent*. In system safety engineering, safety must center on the lack of harm to others that may arise due to the intent itself, or failures arising in an attempt to meet said intent (e.g., implementation failures). The definition of alignment used across AI literature is thus not sufficient to adequately address safety and harms posed by AI-based systems, nor should it be considered a subset of safety. Even if the value alignment is specified to prevent harms and biased outputs according to a given cultural context [39], safety processes requiring risk assessments and additional oversight are necessary regardless of how relatively "good" the system intent may be. Undesirable or unexplored safety concerns [4, 44] will still arise from an AI system meeting well-intentioned specifications (as they do with safety-critical systems), especially given their complexity, scale, and unknown failure modes.

In Section 3.3, we describe how safety requirements are derived from risk assessments, and in Figure 1 below we provide an illustrative example of a system requirement for an Automated Driving Systems (ADS) and a safety requirement that may be derived from it.

| System Requirement | Safety Requirement |
|---|---|
| The ADS shall position itself in the appropriate lane to make an approaching necessary turn. | The ADS shall not cause the vehicle to cross a lane boundary if it would cause any other road user travelling at legal speeds to accelerate, decelerate or corner at more than 2 meters per seconds squared. |

**Figure 1:** Demonstration of a system requirement, with an associated safety requirement to prevent risk of collision

The precise definitions of these terms also impacts the efficacy of techniques adopted from hardware, cybersecurity, and system safety engineering, as these methodologies were created with the distinction between requirements and safety, and the definitions discussed further below, in mind. In the remainder of this paper, we use the term "safety" as intended by system safety engineering.

## 2.2 On Risk Terminology

The terms "hazards," "risks," and "threats" additionally appear in various AI literature, and are often used interchangeably despite having different connotations in the fields from which they are derived.

- **Hazards** are conditions that can result in a system producing harm or undesirable effects to health, life, assets, or the environment.
- **Risks** are assessed within the context of the probability and severity of the hazard becoming reality.
- **Threats** are more specific to the security domain, where an undesirable event can affect the confidentiality, integrity, or availability of the system under consideration. In practice, threats can lead to or be considered as a type of hazard, especially when carrying out security-informed-safety assessments [36].

Although details regarding risk assessments are described further in Section 4.2, we provide an informal illustrative example derived from [20] to demonstrate the distinction between hazards and risks in Figure 2 below. A detailed example can be found in Figure 4.

| Hazard | Risk | |
|---|---|---|
| | Severity | Likelihood |
| Language model generates completions that encode bias in ways that disproportionately harm or benefit different groups. | Critical - Incitement, manipulation, radicalization, or discriminatory harm that may result in physical or mental injuries to multiple people. Cause of consequential error to many individuals. | Occasional (> $10^{-3}$ to < $10^{-1}$) or will occur several times. |

**Figure 2:** Demonstration of hazard and its associated risk through severity and likelihood

Discussions regarding why hazards, and not accidents or failures, are used to assess risk can be found in [24] and provides further context as to why safety is typically not defined as the prevention of failures due to accidents, as is done in [3, 35]. The relationship between individual failures and accidents is not obvious and may be difficult to determine.

We note that risk management frameworks (e.g., ISO/IEC DIS 42001, ISO 9001:2015) are distinct from risk assessments, yet they are often conflated. Risk management is a continuous process to help an organization identify and manage all potential organizational risks. Risk assessments are systematic methodologies that identify, evaluate, and report system risks (e.g., NIST SP 800-30), and are a component of a larger risk management system. Risk management frameworks are beyond the scope of this paper.

## 2.3 On Faults, Failures, and Failure Modes

For completeness and clarity, we additionally define the terms “failure,” ”error,” and “fault,” as they are central to risk assessment frameworks. These terms do not have consistent meanings across standards and domains; here, we follow the common use as defined by ISO/IEC/IEEE 24765:2017.

- **Errors** are erroneous states of the system or human actions that produce an incorrect result.
- **Faults** are a manifestation of an error in a software system.
- **Failures** are a termination of the ability of a system to perform a required function or its inability to perform within specified limits. A failure can be produced when a fault is encountered.
- **Failure modes** are a function manifestation of a failure.

We will use the definitions mentioned above in the remainder of this paper, and we similarly hope that those looking to evaluate safety and risk for AI-based systems will adopt the intended definitions to ensure the integrity of their analyses.

# 3. Pitfalls in Existing Adoptions and Approaches

Various AI-related works have progressively looked to well-established safety and security techniques in an attempt to apply the same methodologies to ML models and their datasets [13, 26, 31, 33, 34, 45, 46]. Adoptions of safety-related techniques, such as Failure Modes and Effects Analysis (FMEA), bug bounties, security threat modeling (i.e., DREAD), and red teaming have been proposed. Unfortunately, due to silos between communities and implicit knowledge held by safety and security practitioners, nuances regarding fundamental issues that limit the application of safety and security techniques to explore AI risks have been lost. This includes the context and processes where specific techniques are intended to be applied within a system life cycle, or the misconstruing of scope and output of these methodologies. Additionally, these works do not provide sufficient detail to fully operationalize the proposed risk modeling approaches to construct comprehensive assurance claims regarding an AI-based system.

Other works have aimed to collate all possible hazards and harms posed by ML models [44]. However, such works do not facilitate or capture how these harms apply to specific domains or systems, their consequential effects, their probability of occurrence (e.g., risk), or how one can operationalize processes for finding the listed harms.

In this section, we outline the pitfalls of referenced or adapted techniques from hardware safety, security, and systems and software safety engineering, including how their limitations and intended applications impact their suitability for use in AI-based systems. Overall, most of these works often point to the adoption of specific techniques without taking into account the larger context and processes in which they are intended to be deployed, or the type of risks they are designed to address.

## 3.1 Limitations and Use of Hardware Safety Techniques

Works such as those in [34] have explored the use of risk modeling from safety-critical domains but neglected to differentiate that *hardware safety* risk techniques (e.g., FMEA) differ from those of general system safety and software engineering techniques—for good reason. A major limitation in adapting hardware techniques is that although hardware behavior is deterministic, hardware failures are based on measured random failures of a hardware component (e.g., malfunction with electromechanical components due to radiation). That is, we have clear expectations regarding the functions of parts, and have an understanding of the failure rates metrics (i.e., Mean Time Between Failures, Mean Time To Failure) and the average repair time of hardware components. Additionally, the emphasis on parts failure entails that emergent system-level failures are not accounted for. Analysis of system behavior and failure modes are thus not within the scope of techniques such as FMEA.

Contrarily, the software safety community itself has recognized that software failures are systematic and not random (e.g., a race condition due to two threads wrongly accessing a shared variable at the same time) [21]. ISO 26262:2018 defines software systematic failures to have a "certain cause that can only be eliminated by a change of the design, … documentation or other relevant factors." Generally, the systematic nature of software failures is directly linked to the intended system behavior and functionality [7, 24]. Attempting to measure the safety properties of software components through random failures is not only technically infeasible, but also not conducive to uncovering the systemic design issues that directly lead to systematic failures. Moreover, the increasing scale and complexity of software[2] has led to software systems seemingly behaving in non-deterministic ways, requiring different approaches from those used to analyze deterministic and bounded hardware behaviors. This is further discussed in the next section.

The issues of applying hardware safety techniques to ML models thus suffer from similar and even further limitations than those for software. Works such as [12] address in great detail why component reliability is insufficient for AI safety, even if applied to AI sub-components. The non-deterministic behavior of ML models and their ever-increasing scale and complexity make such hardware techniques inapplicable to measure the functionality, dependability, and performance of these systems, let alone their use to uncover novel safety harms and ethical implications that are systematic and emergent in nature.

Hardware safety risk techniques such as FMEA aim to target only component-level defects, and they are intended to be deployed within a more general risk framework that subsumes system behavior. More general systems engineering and software risk assessment frameworks [42, 20] have an expanded scope that aims to address hazards, harms, and systematic considerations crucial to software or AI, such as general system failures and emergent behaviors. We discuss such techniques in Section 3.3 and in further detail in Section 4.

## 3.2 Limitations and Use of Cybersecurity Techniques

A key point that the AI community has overlooked pertains to why safety and security communities take differing and sometimes opposing approaches in their risk assessments or threat modeling, respectively. The aim of safety is to prevent a system from impacting its environment in an undesirable or harmful way, typically aiming to protect human lives, the natural environment, or assets by assessing the functionality, performance, dependability, and operability of a system. The aim of security, on the other hand, is to prevent often adversarial environmental agents or conditions from impacting a system in an undesirable or harmful way, aiming to protect the confidentiality, integrity and availability of the information system.

[2] It's very likely that those looking to adapt hardware techniques may have been additionally misled by terms in the literature such as Software FMEA (i.e., SFMEA or SFMECA) that are typically specific to Programmable Logic Controllers (PLCs) and Field Programmable Gate Arrays (FPGAs), which are often too low-level to be considered as such among software and AI developers.

Works such as [13, 31, 46] have looked at adapting cybersecurity practices to uncover hazards or harms for AI-based systems. AI bug bounty programs have been proposed [46] as an attempt to identify and rank risks of harms (rather than vulnerabilities) through security-like threat scoring techniques such as DREAD [38]. However, security threat modeling in general is not suited for measuring hazards and potential harms imposed by AI-based systems. Safety risk frameworks are more appropriate for exploring harms posed by a system (e.g., ML models), rather than threat modeling, which aims to protect a system from its external environment. Take, for example, the use of risk measures adopted from threat scoring. There is a variety of literature noting that security threat scoring frameworks are not developed with academic rigor and yield risk scores that are subjective [19, 38]. It is also important to note that scoring frameworks such as DREAD and Common Vulnerability Scoring System (CVSS) aim to measure the severity and likelihood of vulnerabilities, rather than model the threats of a system on the design level, i.e., at a systematic level.

That is not to say that we should not consider security threat modeling techniques, as they can indeed prevent adversarial cases that compromise the confidentiality, integrity, and availability of a system and can further allow the manipulation of ML models to cause harm. There are also overlapping properties between safety and security, such as the consideration of privacy issues associated with the collection and use of human data and the lack of consent for the subjects of ML models [6, 43]. Works such as those in [30] appropriately articulate a threat model for ML that considers privacy within an adversarial framework. However, if safety and harms are the main concerns, safety risk frameworks can often subsume threat models—recall that a threat can be considered as a type of hazard. Security threat modeling can be deployed in the context of a larger safety risk assessment framework, where they can in turn provide feedback and lead to safety harms (i.e., security-informed-safety [36]).

Other works use the term “red teaming” in the context of large language models (LLMs). In cybersecurity, the intent of a red teaming exercise is to realistically test an organization’s capability to detect and respond to a staged adversarial attack, and to assess and validate security posture and attack resilience. However, the term in AI has come to refer to probing an LLM for harmful outputs, only to update the model latently [13, 31, 29]. These works do not measure the readiness of an LLM to actively combat adversarial inputs (i.e., security), but rather explore potential harms posed by the model (i.e., safety). The likely appropriate technique intended for use in [13, 31] is boundary or stress testing, a verification technique that aims to test edge-cases or fringe inputs that may lead to unknown failure modes and potential hazards.

Understanding the appropriate terminology and frameworks from which we derive techniques is not an issue of pedantry, but allows those seeking to adopt safety and security techniques to understand the taxonomy and existing methodologies that may better aid them in the validation and verification of LLMs or general multi-modal models. Furthermore, misuse of terminology entailing compliance with established safety and security properties, without actually achieving the rigor implied by them, can mislead stakeholders with regard to the claims an AI system satisfies and may provide a false sense of assurance and safety.

## 3.3 Limitations and Use of System Safety Engineering and Software Safety

As previously noted, traditional software failures are systematic, and thus cannot be measured at random. Furthermore, the scope of failure modes possible in software systems differs from that in hardware (i.e., random versus systematic). Unfortunately, there has been little work [12, 20, 47] within the AI community to adapt more general techniques from either the system safety engineering or the software safety community, despite the complexity of these systems being more similar to that of AI. In this section, we look to processes and techniques used within the system safety engineering and software safety community that can be leveraged and adapted to promote safe AI deployment and risk assessment.

Within a safety-critical system, safety-related sub-systems are designed to reduce the frequency (or probability) of hazardous events that may arise when a system meets its specifications or executes its intended functionality. To identify these system hazards, it is necessary to carry out a hazard and risk analysis on hazards or harms, their prevention or mitigation, and performance criteria that define the tolerable risk allowed. For the hazards that require risk reduction measures, a safety function must then be created to meet a specified target Safety Integrity Level (SIL)[3] for implementation[4]. SIL is a measure of system safety performance, in terms of probability of failure on demand (pfd). The safety functions must then meet their target SIL, typically based on an established standard.

Contrary to the standard hardware safety practice of using measured random failures to determine the safety of components, the established solution for software systems is to intentionally build robust software systems by applying increasingly rigorous techniques to meet a SIL to reduce the risk factor of a probability of a failure. This is a concept typically known as “Production Excellence” (PE)[5], where software developers aim to satisfy and certify against SIL criteria outlined in standards such as IEC 61508:2010, DO-178C, and ISO 26262:2018. We note that a hazard analysis and PE are not the only risk exploration or risk reduction measures in a safety assessment. Other techniques, such as System-Theoretic Process Analysis (STPA) [24] or safety-cases [7], are additionally necessary to analyze

[3] Terminology for SIL may differ according to the industry, and include ASIL and DAL.

[4] For control and instrumentation systems, this is where techniques such as Fault Tree Analysis (FTA) and FMEA are deployed to determine the SIL.

[5] Production Excellence is also applicable to hardware and Control & Instrumentation systems.

emergent behavior and the safety properties that arise from complex sub-system interactions.

Despite the well-established effectiveness of safety methodologies and risk assessment approaches for software safety-critical systems, there are two major limitations when lifting these techniques to AI-based systems:

1. It is challenging to identify and determine the tolerable risk for a specific hazardous event or harm within a risk assessment, as the scale and stochastic nature of ML models means that novel and nondeterministic failures (including ethical failures) for AI-based systems cannot be quantified as they have been for control and instrumentation systems, or in software-based practices such as site reliability engineering (SRE).

2. Many of the techniques that allow us to build confidence in the robustness of the system (e.g., static analysis, formal verification, comprehensive testing) are not applicable or transferable to the analysis of ML models, with equivalent techniques yet to be developed or still underway [9].

Even if sufficient formal techniques existed, the use of such methods is based on the assumption that we can determine the functional properties of a system by the way we design and implement it. Only following rigorous recommended techniques, such as those provided in [22] or IEC 61508:2010, will not adequately eliminate the probability of failures, as is the case for even the most complex software systems; since the design of an ML model determines only how it learns, but not what it will learn (i.e., its behavior).

Again, this is not to say that applying such techniques to software infrastructure is not important; in fact, research has shown that systematic software failures of AI infrastructure can propagate and affect the functionality and performance of ML models (such as where a Not a Number (NaN) code error caused uncontrolled acceleration) [17, 41]. However, formal methods must expand beyond system design to data lineage to appropriately address the behavior of ML models. Here, proposed techniques such as [14, 18, 26] would be most applicable to extend the concept of PE to help reduce the risk factor of harms explored. That is, the intention of PE is to help alleviate risk, rather than identify it.

Works such as [20] adapted the more relevant system-level risk assessment framework MIL-STD-882e [42], accompanied by a newly defined set of Hazard Severity Categories (HSC), harms, and losses to accommodate novel harms associated with the use of LLM APIs. Unfortunately, details in [20] on how to operationalize such a risk assessment across a multitude of applications are unclear, especially with consideration of general multi-modal models such as GPT-3, Claude, LaMDA, Bard, and Stable Diffusion. Although a list of “hazard sources” (e.g., Application, System Design, Regulatory and Legal Oversight) is noted, the sources are too general to derive a systematic approach to identify hazards and harms.

In the next section, we outline a blueprint toward comprehensive risk assessments and assurance of AI-based systems that is intended to mitigate against some of the limitations discussed—in particular, to operationalize risk modeling to construct comprehensive assurance claims regarding an AI-based system. This will guide the appropriate determination of the criticality and harms posed by AI-based systems, which can then allow for effective application of PE and safety and security techniques.

# 4. Unifying Risk Assessment and Safety Justification

In this section, we propose a novel systematic risk assessment approach, similar to that described in Section 3.3, but adapted for AI-based systems. We propose a system-level risk assessment to not only define criteria to help determine the tolerable risk allowed, but also to guide system design in order to reduce the frequency of the hazards and harms identified. Our work aims to help developers and auditors build confidence that an AI-based system has adequately addressed its safety risks in its implementation and deployment, so far as is reasonably practicable [2].

## 4.1 AI Operational Design Domain

We propose the integration of the concept of an Operational Design Domain [1], initially introduced for ADS, into risk assessments for more general AI-based systems (e.g., multi-modal models) (see Figure 3). The purpose of an ODD is to provide a "description of the specific operating domain(s) in which an automated function or system is designed to properly operate." Baseline ODDs and scenario analyses are used to identify important functional and safety capabilities. Despite its success within the field of autonomous vehicles, defining an operational envelope to better assess potential risk and required safety functionality has not been at the forefront of the AI algorithmic audit and assurance communities [9, 45]. Indeed, although ADS are AI-based systems, ODDs have not been generalized or extended beyond automotive specific attributes (e.g., physical infrastructure, environmental conditions) to be applicable to a more general class of AI systems.

The majority of algorithmic assessments aim to audit general properties of a system without considering its operational envelope. Although this may be self-evident for some use cases of AI-based systems, the lack of ODD for general multi-modal models has rendered the evaluation of their risk and safety intractable due to the sheer number of applications and therefore, risks posed. This lack of an ODD also prevents those designing or deploying an AI-based system from understanding the constraints under which the system no longer behaves as intended or can escape its designated safety envelope. For example, an ADS which can be safely deployed on a highway in clear weather does not mean it can be deployed on city roads or in adverse weather conditions.

Consider type certification in aviation; certification and risk assessments are carried out for the approval of a particular vehicle design under specific airworthiness requirements (e.g., Federal Aviation Administration 14 CFR part 21). There is no standard assurance or assessment approach for "generic" vehicle types across all domains. It would be contrary to established safety practices and unproductive to presume that the formidable challenge of evaluating general multi-modal models for all conceivable tasks must be addressed first.

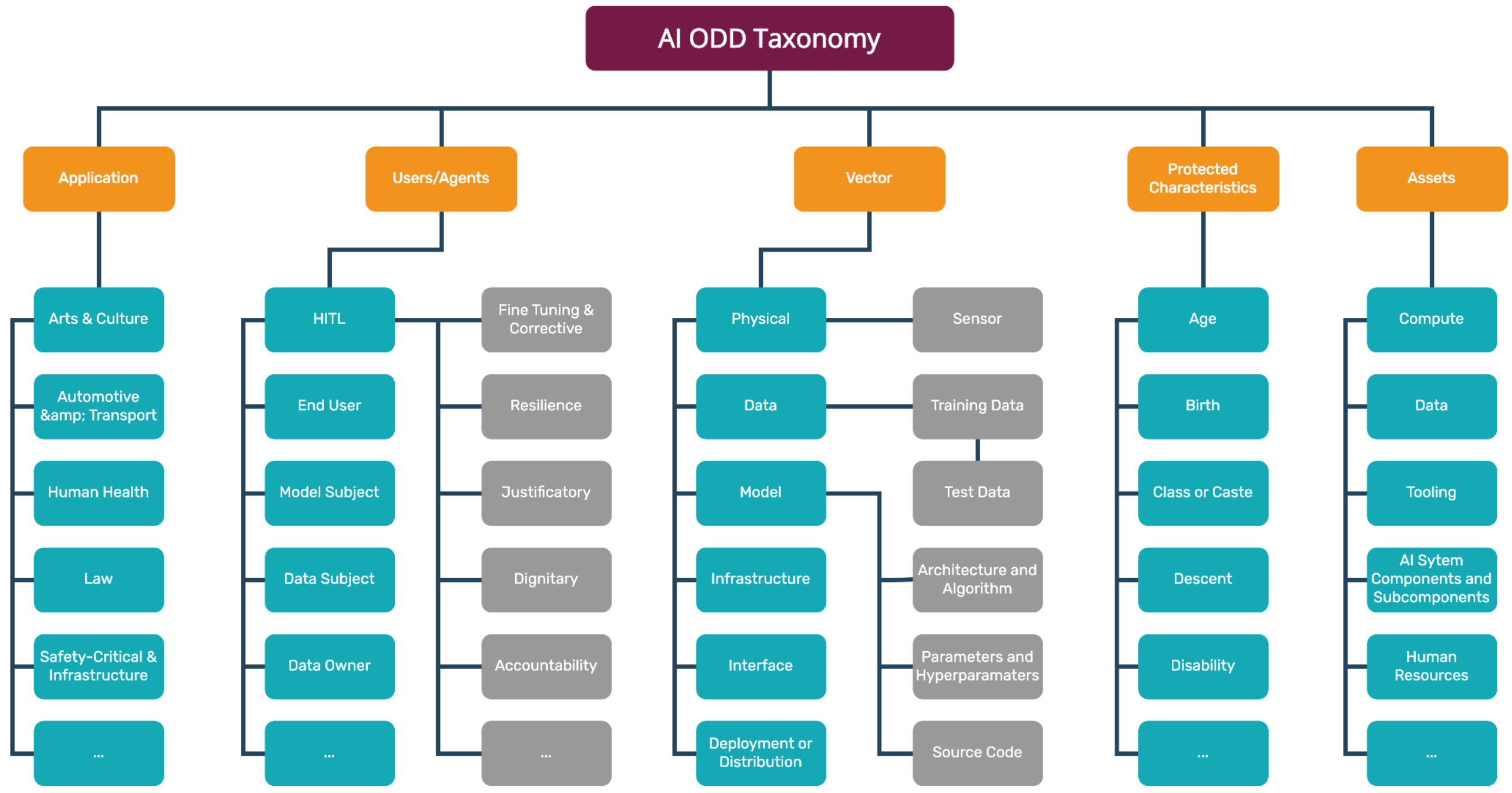

**Figure 3:** AI ODD Taxonomy with the baseline categories and sample subcategories for illustrative purposes

ODDs are intended to be used across an entire system's life cycle, but in this work we focus on their integration into risk assessments, as many stakeholders using AI technologies may not be involved in their development process (i.e., external users). Note that [1] do not provide guidance on how ODDs can be integrated into risk frameworks. To integrate ODDs into a risk framework, it is necessary to define a novel ODD taxonomy relevant to the use of AI technologies, including general multi-modal models. We define a baseline taxonomy below that is subdivided into categories and subcategories, as is done in [1]. We provide justifications or definitions for each category (and subcategories where appropriate). This taxonomy provides a baseline to account for variations of operational envelopes where an AI-based system is used or deployed, and can be built on with further categories or subcategories that may be suitable for the system or organization. A complete ODD taxonomy will include the following:

- **Application/Domain**: Given the applicability of AI systems across all application domains, it is important to be able to map risks against specific applications in order to better determine high-stakes areas or potential harmful impacts. We provide examples of subcategories below, but the area should ultimately be defined to best reflect where a system is deployed for an application domain. For general multi-modal models to be deployed across many domains, a risk assessment must be carried out for each intended use (i.e., per application). Even if risks are applicable across multiple domains, the source and consequences of a potential harm can be reflected differently across each application.

- Arts & Culture
    - Automotive & Transportation
    - Communications
    - Conservation
    - Finance & Economics
    - Government & Civics
    - Human Health
    - Journalism and Media
    - Law
    - Manufacturing and Commerce
    - Marketing, Advertising, and Microtargeting
    - Opportunity and Livelihood
    - Productivity and Education
    - Safety-Critical and Infrastructure
    - Security and Defence
    - Science, Technology, Engineering, and Mathematics
    - Social and Political Advocacy

- **Users/Agents:** Users or human agents within an AI system life cycle have taken on varied and complex roles, often with an unclear purpose or goal. For example, the use of humans for corrective measures (i.e., fine-tuning, data labeling) has been viewed as a way to shift accountability to a "Human In The Loop" (HITL) and to claim that an AI-system is inherently value-aligned as a means to overlook further safety measures [28] (recall the discussion in Section 2.1). However, as shown in [10], human bias infiltrates all aspects of data use, and regardless of where a HITL is placed within an AI system life cycle, the lack of detail regarding the role a human is expected to play obscures the risks their use (or lack thereof) may bring about. [11] lays out a comprehensive typology of possible roles intended to clarify the purpose and intended output of a human. We consider these defined roles below for our ODD under "HITL" given that human factors integration has always been a pillar of the safety of complex critical systems. System safety engineering has always recognized the importance of human-machine interactions to the overall safety and functionality of the system.

  Aside from HITL, there are numerous risks to privacy and confidentiality in AI-based systems that have long been discussed. Ongoing scrutiny has particularly been associated with the collection and use of human data [30], the potential risks for bias and discrimination, and lack of consent for the subjects of ML models [6, 43]. Therefore, we include these human subjects in our ODD taxonomy, in addition to agents that develop or deploy AI-based systems, given the importance of understanding hazards and harms beyond model performance.

- End user
- Model user
- Model subject
- Data subject
- Data owner
- Model owner
- Data Labeler - Training
- Human In The Loop (HITL) [11]
  - Fine Tuning & Corrective
  - Resilience
  - Justificatory
  - Dignitary
  - Accountability
  - Stand in
  - Friction
  - Warm body
  - Interface link

- **Vector:** The vector or attack surface is an important attribute to consider when assessing the scope and mitigations for hazards or harms. Additionally, a system's interfaces and how users or agents interact with it, implicitly or explicitly, can help identify the reach or cascading effects of hazards or harms derived from a specific interface vector. We build on the vector or surface introduced in [30], where systems using ML models are viewed as a generalized data processing pipeline. This includes consideration of how features are collected from data repositories, how data is processed in the digital domain and used by a model to produce an output, and how the output is communicated to an external system or user and acted upon. This is particularly relevant for assessments analyzing function and system-level component hazards.
  - Physical or Physical-Infrastructure
    - Sensor
    - Cyber-physical
    - Corporeal
  - Data
    - Training data
    - Test data
    - Input data
  - Model

- Architecture and algorithm
        - Parameters and Hyperparameters
        - Source Code
        - Trained model representation
    - Software Infrastructure
    - Interface
        - UI/UX
        - APIs
        - I/O
    - Deployment and Distribution
        - Management
        - Monitoring
        - Maintenance

- **Protected Characteristics [16, 28, 32]:** Given that the majority of harms identified by the deployment and use of ML models disproportionately affect protected groups [4, 6], it is necessary to consider these harms in every aspect of an AI system's life cycle. These characteristics should be expanded to consider further geographical and cultural contexts (see discussion in [39]).

    - Age
    - Birth
    - Class or Caste
    - Descent
    - Disability
    - Gender identity
    - Genetic information
    - Health status
    - Language
    - Marriage and civil partnership
    - Migration status
    - National, ethnic, or social origin
    - Political/other opinion
    - Pregnancy and maternity
    - Property, birth, other status
    - Race
    - Religion or belief
    - Sex
    - Sexual Orientation

- **Assets**: As discussed in Section 3.2, cybersecurity risks are still prevalent and may impact the safety of a system. It is important for an organization to account for the confidentiality, privacy, integrity, and availability of its assets in order to fully understand and address the risks and impacts for at least the following:
    - Compute
    - Data
    - Tooling
    - AI system components and subcomponents
    - Human Resources
    - Monetary
    - Physical

Note that there is overlap between the subcategories of each category (e.g., Data). The overlap between such subcategories should be considered given the definition and the intention of the category that subsumes it. We encourage those deploying AI-based systems to consider defining further categories and subcategories beyond these baseline ODDs where relevant to their own practice domain (e.g., [1]). For example, an ODD may optionally include value-alignment properties, such as those noted in Sensitive Topics or Sentiment Positions for Social Context in [39]. Recall that many of the attributes defined in value-alignment literature are more suitable for system requirements. We believe that further refined operational domains will allow stakeholders to identify more granular risks relevant to their application and use of AI-based systems. We refer to [1] for examples of how to further refine categories with an ODD taxonomy.

## 4.2 Operationalizing Risk Assessments

A common challenge encountered when performing a hazard analysis is comprehensively identifying hazards given consideration of all possible and relevant scenarios [25]. In this section, we demonstrate how an ODD can be used to systematically identify risks once the relevant taxonomy has been defined or identified. The defined ODD categories and subcategories must be considered, even if they are not perceived as risks. The goal of the risk assessment is then to identify the risk within a defined operational envelope or scenarios.

We build on the risk framework defined in [20] given its novel definitions of HSCs, harms, and losses tailored for the use of general multi-modal models. Hazards are, by definition, linked to specific types of harms or losses that stakeholders identify. As in traditional system risk assessments, we use a Hazard Risk Index (HRI) as a metric to note the risk for each hazard. As noted in Section 2.2, an HRI (i.e., risk) is based on the product of the probability of a hazard condition against its severity. Quantitative data or a quantitative probability guide with corresponding qualitative metrics (i.e., Frequent, Probable, Occasional, Remote, Improbable) can be used. The HSCs, losses, and HRI utilized are available in Appendix A for reference. Recall that both determining and mitigating for probabilities of failures in AI-based systems is an open problem, since the design of an ML model determines only how it learns, not what it learns (i.e., its behavior). The determination of a likelihood of a hazard or harm is beyond the scope of this paper, and requires the development of sophisticated techniques from machine learning to social science to guide the development of adequate risk metrics relevant to specific subject domains [5, 9, 27].

Risk assessments are carried out at various levels of abstraction, such as product specifications, architecture and design, implementation, hardware components (e.g., FMEA), and operation and maintenance. The assessments are typically split among multidisciplinary teams with different backgrounds such as policy, safety, security, engineering, and law, to ensure comprehensive coverage of potential harms and hazards. We refer to the list of "hazard sources" in [20] (e.g., Application, System Design, Regulatory—as a preliminary breakdown. As noted in Section 3.3, hazard analyses should not be used as the sole technique for risk exploration. For AI-based systems, complementary techniques equivalent to "Production Excellence" such as [14, 18, 22], should also be considered or integrated into the risk assessment process.

Given the above, the ODD can be seen as a consideration of various permutations of the categories to effectively explore a  wide range of scenarios and their associated risks. With an identified application domain and a system abstraction scope defined for the risk assessment, assessors should enumerate through all categories and subcategories and consider all hazards that may arise due to their interaction. This can be best illustrated with the example figure below, where each hazard has an associated field for each ODD category. We provide three distinct examples of hazard entries across varying application domains and levels of abstraction of the system.

| Hazard Source | Hazard Description | Trigger Event | Application/Domain | Users/Agents | Vector | Protected Characteristic | Assets | Potential Effects |
|---|---|---|---|---|---|---|---|---|
| System Design - Model Training | Subpopulations not appropriately identified or represented within distributions of the dataset. | Use of CNN skin cancer classifier on skin on a non-white skintone. | Human Health | End user | Data - Training | Race | Data | Increased risk of false negative, or misclassification of skin cancer as a benign or not present. |
| | | | | | | | | |
| System Implementation | Not a Number (NaN) value propagates to speed limit value in autonomous vehicle planning module. | Lack of float precision and calculation leading to Not a Number (NaN). | Automotive & Transportation | Model user | Model - Source Code | N/A | AI system components and subcomponents | NaN causing uncontrolled acceleration leading to increased risk of physical harm, injury, or death. |
| | | | | | | | | |
| Human Rights | Exposure to toxic datasets or model inputs that promote hate crimes towards a labeler's gender identity. | Labeling and fine-tuning datasets for violence, hate speech, etc., to train model to learn to detect those forms of toxicity. | Marketing, Advertising, and Microtargeting | HITL - Fine Tuning & Corrective | Data - Training | Gender identity | Human Resources | Mental distress, grief, and potential impact to long-term mental illnesses (e.g., PTSD). |

**Figure 4:** Demonstration of hazard entries within a hazard analysis using ODD categories (colored) as fields for consideration

Note that in Figure 4, the "HRI" and "Mitigation" fields have been omitted for presentation purposes. The HRI for each hazard assists in understanding how the risks compare to each other, and the priority with which each hazard must be controlled. "Mitigations" describe specific actions that would reduce the associated HRI for a hazard. Action points arising from mitigations comprise the associated system safety requirements (e.g., Figure 1) that accompany system or operational requirements. With each set of mitigations implemented, the HRIs should be recalculated, and the process should be repeated until all undesirable risk is eliminated for the safe deployment of an AI-based system. A risk template for use with the corresponding ODD categories is available in Appendix A, Table 3.

# 5. Conclusive Remarks

In order to build and deploy safe AI-based systems, it is important to align on key terminology and techniques in order to level the playing field on assurance objectives and how they can be achieved. For example, a lack of cohesion on baseline terminology such as the distinction between "safety" and "alignment" has led to contradictory approaches that incorrectly equate safety measures with requirements. This can lead to drastically different outcomes when constructing risk assessments, and misaligned principles and goals. We outline a consistent and comprehensive terminology and discuss fundamental issues with the adoption of hardware, cybersecurity, and system safety techniques that limit their direct application to AI risk assessments.

Finally, we propose a novel assurance approach that overcomes some of the discussed limitations for AI risk assessment through the use of ODDs, a concept initially introduced for automated driving systems. To integrate ODDs into a risk framework, we define a novel ODD taxonomy relevant to the use of AI technologies, including general multi-modal models. The taxonomy is subdivided into categories and subcategories, with categories including Application/Domain, Users/Agents, Vector, Protected Characteristics, and Assets. These categories allow for the exploration of a wide range of scenarios and their associated risks. We believe that by defining a more concrete operational envelope, developers and auditors can better assess potential risks and required safety mitigations for AI-based systems.

# Appendix A

| Description | Category | Definition (Mapped to Table 2) |
|---|---|---|
| Catastrophic | 1 | Death, permanent total disability, direct harm, significant system or asset loss, or irreversible significant environmental impact. |
| Critical | 2 | Incitement, manipulation, radicalization, or discrimination that may result in mental or physical harm. Cause of consequential error to many individuals, reversible significant environmental impact, or considerate asset loss. |
| Major | 3 | Injury or cause of consequential error to a few individuals, reversible environmental impact, or moderate asset loss. |
| Minor | 4 | Injury or cause of consequential error not resulting in any long term harm, minimal environmental impact, or negligible asset loss. |

**Table 1:** Hazard Severity Categories associated with the use of general multi-modal models [20 , 42]

| Hazard Frequency | Catastrophic | Critical | Marginal | Negligible |
|---|---|---|---|---|
| **(A) Frequent** | **1A** | **2A** | **3A** | **4A** |
| **(B) Probable** | **1B** | **2B** | **3B** | **4B** |
| **(C) Occasional** | **1C** | **2C** | **3C** | **4C** |
| **(D) Remote** | **1D** | **2D** | **3D** | **4D** |
| **(E) Improbable** | **1E** | **2E** | **3E** | **4E** |

**Table 2:** Hazard Risk Index [42] considering hazard frequency against its severity category

| ID | Hazard Source | Hazard Description | Trigger Event | Application/Domain | Users/Agents | Vector | Protected Characteristics | Assets | Potential Effects | HRI | Recommend Mitigations | HRI (post) |
|---|---|---|---|---|---|---|---|---|---|---|---|---|
| H1 | System Design - Model Training | Subpopulations not appropriately identified or represented within distributions of the dataset. | Use of CNN skin cancer classifier on skin on a non-white skin tone. | Human Health | End user | Data - Training | Race | Data | Increased risk of false negative, or misclassification of skin cancer as a benign or not present. | 2B | | |
| | | | | | | | | | | | | |
| | | | | | | | | | | | | |
| | | | | | | | | | | | | |
| | | | | | | | | | | | | |
| | | | | | | | | | | | | |

**Table 3:** Risk assessment template with ODD categories as fields